\documentclass[aps,showpacs,showkeys,preprint]{revtex4}
\usepackage[dvips]{graphicx}
\usepackage{latexsym}

\begin{document}

\preprint{\vbox{\hbox{\tt gr-qc/0703019}}}

\title{Transition from AdS universe to DS universe in the BPP model}

\author{Wontae Kim}
  \email{wtkim@sogang.ac.kr}
  \affiliation{Department of Physics and Center for Quantum Spacetime,
    Sogang University, C.P.O. Box 1142, Seoul 100-611, Korea}
\author{Myungseok Yoon}
  \email{younms@sogang.ac.kr}
  \affiliation{Center for Quantum Spacetime, Sogang
    University, Seoul 121-742, Korea}

\date{\today}

\begin{abstract}
It can be shown that in the BPP model the smooth phase transition from
the asymptotically decelerated AdS universe to the asymptotically
accelerated DS universe is possible by solving the modified
semiclassical equations of motion. This transition comes from
noncommutative Poisson algebra, which gives the constant curvature
scalars asymptotically. The decelerated expansion of the early
universe is due to the negative energy density with the negative
pressure induced by quantum back reaction, and the accelerated
late-time universe comes from the positive energy and the negative
pressure which behave like dark energy source in recent cosmological
models.
\end{abstract}


\maketitle

\section{Introduction}
\label{sec:intro}
One of the intriguing issues is not only to describe the late-time
accelerated expansion of our universe but also to explain the smooth
transition from decelerating phase to accelerating one. In the context
of Einstein theory of general relativity, the accelerating universe
means that the parameter $\omega\equiv p/\varepsilon$ of equation of
state \cite{carroll} is negative, where $\varepsilon$ and $p$ are the
energy density and the pressure, respectively. So, in the ordinary
Friedmann equation, the energy density is assumed to be positive while
the pressure is negative. Even more $\omega<-1$ can be required to
compensate the effect of ordinary matters in our universe. In some
sense, it implies that the state parameter may depend on time and make
it possible to explain the transition from decelerating phase to
accelerating one. In the quintessence model based on supergravity or
M/string theory, the transition has been studied in terms of the
numerical simulation \cite{gardner}. 

On the other hand, a two-dimensional dilaton gravity may be useful in
studying the transition from the decelerated phase and the accelerated
phase because there are fewer degrees of freedom rather than the
four-dimensional counterpart. Furthermore, there exist exactly soluble
models semiclassically \cite{cghs,bc,rst,bpp,bk,kv}, whose quantum
back reactions of the geometry are easily treated so that various 
cosmological problems have been studied in
Refs. \cite{bk,ky:dg,rey,gv,  ky:bd,cdkz}. However, in even this
semiclassically soluble gravity, it is difficult to realize the smooth
phase transition because the solution shows the only decelerating or
accelerating behavior. Recently, it has been shown that it is possible
to obtain the transition from decelerating phase to accelerating one
by assuming the modified Poisson brackets \cite{ky:au} corresponding
to noncommutativity of fields \cite{bn,vassilevich,ko}. Unfortunately,
the future singularity appears at finite time in this model, and the
decelerated geometry has been patched by hand for the regularity. 

So, in this paper, we would like to study the smooth phase transition
from the decelerated expansion to accelerated expansion without any
curvature singularity in the Bose-Parker-Peleg (BPP) model \cite{bpp},
which is one of the exactly soluble model semiclassically. In
particular, even though the classical cosmological constant is not
assumed, the initial state is asymptotically anti-de Sitter (AdS) and
the late time behavior of our universe is asymptotically  de Sitter
(DS). This interesting feature is due to the noncommutativity in the
modified Poisson algebra. In Sec.~\ref{sec:bpp}, we find the
semiclassical Hamiltonian in the BPP model and also define semiclassical
energy-momentum tensors, and obtain the energy density and the
pressure in view of a perfect fluid. In Sec.~\ref{sec:pb}, solving the
semiclassical Hamiltonian equations of motion with the ordinary
Poisson brackets in the BPP model, we obtain the accelerated expansion
solution. In Sec.~\ref{sec:mpb}, we will take the modified Poisson
brackets instead of the conventional Poisson algebra. Under some
conditions for integration constants, the solution shows that the
smooth transition from AdS (decelerating) phase at the past infinity
to DS (accelerating) phase is possible. Finally, some discussions are
given in Sec.~\ref{sec:dis}.

\section{Hamiltonian and  energy-momentum tensors } 
\label{sec:bpp}
In the low-energy string theory, the two-dimensional dilaton gravity
are described by
\begin{equation}
  S_{\rm DG} = \frac{1}{2\pi} \int d^2 x \sqrt{-g} e^{-2\phi} \left[ R
    + 4 (\nabla\phi)^2 + 4 \lambda^2 \right],   \label{action:dg} 
\end{equation}
and the conformal matter fields is given as
\begin{equation}
  S_{\rm cl}= \frac{1}{2\pi} \int d^2 x \sqrt{-g} \left[ - \frac12
    \sum_{i=1}^N (\nabla f_i)^2 \right],   \label{action:cl}
\end{equation}
where $\phi$ and $f_i$'s are the dilaton and the conformal matter
fields, respectively. We set the vanishing cosmological constant
$\lambda^2=0$ for simplicity in what follows. The quantum effective
action for the conformal matter~(\ref{action:cl}) is written as
\begin{equation}
  \label{action:qt}
  S_{\rm qt} = \frac{\kappa}{2\pi} \int d^2 x \sqrt{-g} \left[
    -\frac14 R \frac{1}{\Box} R + (\nabla\phi)^2 - \phi R \right],
\end{equation}
where $\kappa=(N-24)/12$. The first term in Eq.~(\ref{action:qt})
comes from the Polyakov effective action of the classical matter
fields~\cite{cghs,rst} and the other two local terms have been
introduced in order to solve the semiclassical equations of motion
exactly~\cite{bpp}. The higher order of quantum correction beyond the
one-loop is negligible in the large $N$ approximation where $N \to
\infty$ and $\hbar \to 0$, so that $\kappa$ is assumed to be positive
finite constant.

In order to study consider the quantum back reaction semiclassically,
we take the total action as
\begin{equation}
  \label{action:total}
  S = S_{\rm DG} + S_{\rm cl} + S_{\rm qt}.
\end{equation}
In the conformal gauge, $ds^2 = -e^{2\rho} dx^+ dx^-$, the total
action and the constraint equations are written as 
\begin{eqnarray}
 S &=& \frac{1}{\pi} \int\/ d^2 x \bigg[ e^{-2\phi}\left(
       2\partial_+\partial_-\rho - 4\partial_+\phi\partial_-\phi
       \right) - \kappa\big( \partial_+\rho\partial_-\rho
       + 2\phi\partial_+\partial_-\rho \nonumber \\
   & & + \partial_+\phi\partial_-\phi \big) + \frac12 \sum_{i=1}^{N}
       \partial_+f_i\partial_-f_i \bigg] \label{action:conf}
\end{eqnarray}
and
\begin{eqnarray}
  & & e^{-2\phi}\left[ 4\partial_\pm\rho\partial_\pm\phi
      - 2\partial_\pm^2\phi \right]
      + \frac12\sum_{i=1}^N\left(\partial_\pm f_i \right)^2
      + \kappa\left[ \partial_\pm^2\rho
      - \left(\partial_\pm\rho\right)^2\right] \nonumber \\
  & & \qquad\qquad\qquad - \kappa\left(
      \partial_\pm^2\phi - 2\partial_\pm\rho\partial_\pm\phi\right)
      - \kappa\left(\partial_\pm\phi\right)^2
      - \kappa t_\pm = 0, \label{constr:conf}
\end{eqnarray}
where $t_{\pm}$ reflects the nonlocality of the induced gravity of the
conformal anomaly. Then, we take the vanishing classical matter, $f_i
= 0$ in order to take into account only the quantum-mechanically
induced source. Defining new fields as~\cite{bc,bpp}
\begin{eqnarray}
\Omega &=& e^{-2\phi}, \label{def:Omega} \\
\chi &=& \kappa (\rho-\phi) + e^{-2\phi}, \label{def:chi} 
\end{eqnarray}
the gauge fixed action is obtained in the simplest form of 
\begin{equation}\label{action:new}
S = \frac{1}{\pi} \int\/d^2 x \left[
    \frac{1}{\kappa}\partial_+\Omega\partial_-\Omega 
    - \frac{1}{\kappa}\partial_+\chi\partial_-\chi \right]
\end{equation}
and the constraints are given by
\begin{equation}
\kappa t_\pm = \frac{1}{\kappa}\left( \partial_\pm\Omega\right)^2 
    - \frac{1}{\kappa} \left(\partial_\pm\chi\right)^2 
    + \partial_\pm^2\chi. \label{constr:new}
\end{equation}

In the homogeneous space, using the relations of $x^\pm = t \pm x$,
the Lagrangian and the constraints are obtained, 
\begin{eqnarray}
  L &=& \frac{1}{2\kappa} \dot{\Omega}^2 
        - \frac{1}{2\kappa} \dot{\chi}^2, \label{L} \\
  \frac{1}{4\kappa} \dot{\Omega}^2 \!&-&\! \frac{1}{4\kappa}
        \dot{\chi}^2 + \frac14 \ddot{\chi} - \kappa t_{\pm}=0, 
        \label{con}
\end{eqnarray}
where the action is redefined by $S/L_0 = \frac{1}{\pi} \int dt L$
with $L_0=\int dx$, and the overdot denotes the derivative with
respect to the conformal time $t$. Then, the Hamiltonian becomes
\begin{equation}
  \label{H}
  H = \frac{\kappa}{2} P_\Omega^2 - \frac{\kappa}{2} P_\chi^2
\end{equation}
in terms of the canonical momenta $P_\chi = - \frac{1}{\kappa}
\dot{\chi}$, $P_\Omega = \frac{1}{\kappa} \dot{\Omega}$.

Since the semiclassical energy-momentum tensors are defined by
$T_{\mu\nu}^{\rm qt} \equiv - (2\pi /\sqrt{-g}) (\delta S_{\rm qt} /
\delta g^{\mu\nu})$, they can be written as 
\begin{eqnarray}
  T_{\pm\pm}^{\rm qt} &=& -\kappa t_\pm + \kappa \partial_\pm^2 (\chi
    - \Omega) - \kappa [\partial_\pm(\chi - \Omega)]^2 \nonumber \\
    &=& -\kappa t_\pm + \frac14 (\ddot\chi - \ddot\Omega) -
    \frac{1}{4\kappa} (\dot\chi - \dot\Omega)^2, \label{T++} \\
  T_{+-}^{\rm qt} &=& -\kappa \partial_+ \partial_- (\chi - \Omega)
    \nonumber \\
    &=& -\frac14 (\ddot\chi - \ddot\Omega). \label{T+-}
\end{eqnarray}
They can be regarded as a perfect fluid written in the form of
\begin{equation}
  \label{def:perfect}
  T_{\mu\nu}^{\rm qt} = p g_{\mu\nu} + (p+\varepsilon) u_\mu u_\nu,
\end{equation}
where $\varepsilon$ and $p$ are the energy density and the pressure,
respectively, and $u^\mu$ is the 4-velocity vector field of flow. In
the comoving coordinate, $ds^2 = -d\tau^2 + a^2(\tau) dx^2$, the
4-velocity is given by $u_\mu = (1,0)$, and then we can obtain the
distributions of the energy density and the pressure. Note that the
comoving time are related to the conformal time, $\tau=\int e^{\rho}
dt = \int \Omega^{-1/2} \exp[(1/\kappa)(\chi-\Omega)] dt$. Then, the
energy density and pressure are written as 
\begin{eqnarray}
  \varepsilon &=& T_{\tau\tau}^{\rm qt} 
    = e^{-2\rho}(T_{++}^{\rm qt} + 2 T_{+-}^{\rm qt} 
      + T_{--}^{\rm qt}), \label{def:energy} \\ 
  p &=& \frac{1}{a^2} T_{xx}^{\rm qt} 
    = e^{-2\rho}(T_{++}^{\rm qt} - 2 T_{+-}^{\rm qt} 
      + T_{--}^{\rm qt}). \label{def:pressure} 
\end{eqnarray}
Note that the state parameter $\omega$ has been defined as the
equation of state $p=\omega \varepsilon$.

\section{Accelerated universe with conventional Poisson brackets}
\label{sec:pb}
In this section, we would like to recapitulate the evolution of the 
two-dimensional universe by solving the semiclassical
equations of motion in the BPP model. Even if the solutions can 
be obtained directly from the Lagrangian equations of motion, we will
solve them in terms of the Hamiltonian formulation since the latter
case is more convenient to modify the original equations of motion. 
Let us now define the conventional Poisson brackets,
\begin{equation}
  \label{PB:com}
  \{\Omega, P_\Omega\}_{\mathrm{PB}} = \{\chi, P_\chi\}_{\mathrm{PB}} =1,
  \quad \mathrm{others} = 0
\end{equation}
and then the Hamiltonian equations of motion in Ref.~\cite{bk:H} are
given by $\dot {\cal O}  = \{ {\cal O}, H \}_{\mathrm{PB}}$ where
${\cal O}$ represents fields and corresponding momenta. Then they are
explicitly written as
\begin{eqnarray}
   & & \dot\chi = - \kappa  P_\chi, \quad \dot\Omega = \kappa
       P_\Omega, \label{1st:x_com} \\
   & & \dot{P}_\chi =0, \quad \dot{P}_\Omega = 0. \label{1st:p_com}
\end{eqnarray}
Since the momenta $P_\Omega$ and $P_\chi$ are constants of motion,
we can easily obtain the solutions,
\begin{eqnarray}
  & & \Omega =  \kappa P_{\Omega_0} t + A_0, 
      \label{sol:Omega_com} \\
  & & \chi =  -\kappa P_{\chi_0} t + B_0,
      \label{sol:chi_com}
\end{eqnarray}
where $P_\Omega = P_{\Omega_0}$, $P_\chi = P_{\chi_0}$, $A_0$, and
$B_0$ are arbitrary constants. From the definition (\ref{def:Omega}),
the solution $\Omega$ in Eq.~(\ref{sol:Omega_com}) must be
positive. This leads to three cases of conformal time $t$: one is $t >
A_0 / (\kappa P_{\Omega_0})$ with $P_{\Omega_0} > 0$, another is $t <
A_0 / (\kappa P_{\Omega_0})$ with $P_{\Omega_0} < 0$, and the other is
$-\infty < t < \infty$ with $P_{\Omega_0} = 0$ and $A_0 > 0$.

Next, the dynamical solutions~(\ref{sol:Omega_com}) and
(\ref{sol:chi_com}) should by satisfied with constraint~(\ref{con}),
which results in 
\begin{equation}
  \label{constr:com}
  \kappa t_\pm = \frac{\kappa}{4} (P_{\Omega_0}^2 - P_{\chi_0}^2).  
\end{equation}
Note that the integration functions $t_\pm$ determined by
the matter state are time-independent. On the other hand, by using
Eqs.~(\ref{sol:Omega_com}) and (\ref{sol:chi_com}), the curvature
scalar is calculated as 
\begin{equation}
  \label{R:com}
  R = \frac{2}{a}\frac{d^2a}{d\tau^2} 
    = \kappa^2 P_{\Omega_0}^2 e^{-2\rho + 4\phi} 
    =  \kappa^2 P_{\Omega_0}^2 \frac{e^{ -2B_0 +2 \kappa P_{\chi_0} 
      t} }{ A_0 + \kappa P_{\Omega_0} t } \ge 0, 
\end{equation}
where the equality corresponds to the case of $P_{\Omega_0}=0$ and
$A_0>0$, in other words, which means flat spacetime. 

Plugging  the constraint~(\ref{constr:com}) into Eqs.~(\ref{T++}) and
(\ref{T+-}), the induced energy-momentum tensors are explicitly
written as
\begin{eqnarray}
  T_{\pm\pm}^{\rm qt} &=& -\frac{\kappa}{2} P_{\Omega_0}(P_{\Omega_0}
    + P_{\chi_0}), \label{T++:com} \\
  T_{+-}^{\rm qt} &=& 0,
\end{eqnarray}
which yields from Eqs.~(\ref{def:energy}) and (\ref{def:pressure}),
\begin{equation}
  \varepsilon = p = -\kappa P_{\Omega_0}(P_{\Omega_0} +
    P_{\chi_0}) (\kappa P_{\Omega_0} t + A_0) \exp\left[ \frac{2}{\kappa}
    (A_0 - B_0) + 2 (P_{\Omega_0} + P_{\chi_0}) t \right].
    \label{energy:com}
\end{equation}
Note that the state parameter is simply $\omega=1$ in this
semiclassical case, and the curvature scalar which is proportional to
the acceleration is always positive under the condition of $\Omega
>0$. So, there is no phase transition from the deceleration to the
acceleration, and we can not obtain the AdS-DS phase transition. 


\section{AdS-DS phase transition with modified Poisson brackets}
\label{sec:mpb}
In this section, we now study whether the phase change of the universe
is possible or not in the context of the modified semiclassical
equations of motion. The similar analysis to the previous section will
be done along with the noncommutative algebra~\cite{bn,sw}, 
\begin{eqnarray}
  & & \{ \Omega, P_\Omega \}_{\mathrm{MPB}} = 
      \{ \chi, P_\chi \}_{\mathrm{MPB}} = 1, \nonumber \\ 
  & & \{ \chi, \Omega \}_{\mathrm{MPB}} = 0
      , \quad \{ P_\chi, P_\Omega \}_{\mathrm{MPB}} = 
      \theta, \quad \mathrm{others} = 0, \label{PB:non}
\end{eqnarray}
where $\theta$ is a positive constant. Note that our starting
semiclassical action seems to be quantized one more, however, this is
not the case since these modified Poisson brackets are simply the
counterpart of the conventional Poisson brackets which are not quantum
commutators. If the fields had been taken as operators by decomposing
the positive and the negative frequency modes along with the normal
ordering, then it would be the quantization of a quantization. But our
modified Poisson brackets just modify the conventional
(semiclassical) Hamiltonian equations of motion, which still result
in  the semiclassical solutions, of course, they are
$\theta$-dependent due to the modification of the Poisson brackets.   

Using the Hamiltonian~(\ref{H}), the previous
equations of motion are promoted to the followings,
\begin{eqnarray}
  & & \dot\chi = \{ \chi, H \}_{\mathrm{MPB}} = -\kappa P_\chi, \quad
      \dot\Omega =  \{ \Omega, H \}_{\mathrm{MPB}} = \kappa P_\Omega, 
      \label{1st:x_non}\\
  & & \dot{P}_\chi =  \{ P_\chi, H \}_{\mathrm{MPB}} =\kappa \theta
      P_\Omega, \quad \dot{P}_\Omega = \{ P_\Omega, H \}_{\mathrm{MPB}} 
      = \kappa\theta P_\chi. \label{1st:p_non}
\end{eqnarray}
Note that the momenta are no more constants of motion because of
nonvanishing $\theta$, hereby, a new set of equations of motion from
Eqs.~(\ref{1st:x_non}) and (\ref{1st:p_non}) are obtained,
\begin{equation}
  \label{eom:non}
  \ddot\chi = -\kappa \theta \dot\Omega, \quad \ddot\Omega =
  -\kappa \theta \dot\chi.
\end{equation}
Of course, the parameter $\theta$ is independent of the quantization
where the modified semiclassical equations of motion (\ref{1st:p_non})
is reduced to Eq. (\ref{1st:p_com}) for $\theta \rightarrow 0$.
From the coupled equations of motion ~(\ref{eom:non}), we obtained the
solutions as
\begin{eqnarray}
 \Omega &=& e^{-2\phi} = \alpha e^{-\kappa\theta t} + \beta
    e^{\kappa\theta t} + A, \label{nc:Omega} \\
 \chi &=& e^{-2\phi} + \kappa (\rho - \phi)  = \alpha
    e^{-\kappa\theta t} - \beta e^{\kappa\theta t} +
    B, \label{nc:chi}
\end{eqnarray}
where $\kappa$ has been assumed to be a positive constant, and
$\alpha$, $\beta$, $A$, and $B$ are constants of integration. Since
$\Omega$ should be positive in Eq.~(\ref{nc:Omega}), the constants
$\alpha$, $\beta$, and $A$ are appropriately restricted. Then, the
scale factor and the expanding velocity are given as
\begin{eqnarray}
 a(\tau) &=& e^{\rho(t)} = \frac{\exp[- \frac{1}{\kappa} (A-B) -
   \frac{2}{\kappa} \beta e^{\kappa\theta t}]}{\sqrt{\alpha
   e^{-\kappa\theta t} + \beta e^{\kappa\theta t} +
   A}},  \label{nc:a} \\
 \frac{da}{d\tau} &=& \dot\rho = \frac12 \theta  \frac{\kappa(\alpha
   e^{-\kappa\theta t} - \beta e^{\kappa\theta t}) - (2\beta
   e^{\kappa\theta t} + A)^2 + A^2 - 4
   \alpha\beta}{\alpha e^{-\kappa\theta t}
   + \beta e^{\kappa\theta t} + A}, \label{nc:vel} 
\end{eqnarray}
respectively, where we used $dt/d\tau = e^{-\rho(t)}$ and $a(\tau) =
e^{\rho(t)}$. The overdot denotes the derivative with respect to $t$
and comoving time $\tau$ is related to conformal time $t$ by $\tau =
\int e^{\rho(t)} dt$, which can be explicitly calculated from the
scale factor~(\ref{nc:a}). Subsequently, the acceleration and the
curvature scalar are calculated as 
\begin{eqnarray}
 \frac{d^2 a}{d\tau^2} &=& e^{-\rho} \ddot\rho = \frac12 \kappa
    \theta^2 \frac{\exp[ \frac{1}{\kappa} (A - B) +
    \frac{2}{\kappa} \beta e^{\kappa\theta t}]}{\sqrt{\alpha
    e^{-\kappa\theta t} + \beta e^{\kappa\theta t} + A}}
    \bigg[\kappa \frac{A^2 - 4\alpha\beta}{\alpha
    e^{-\kappa\theta t} + \beta e^{\kappa\theta t} + A}
    \nonumber \\ 
 & & \quad\qquad\  - (2\beta e^{\kappa\theta t} + A)^2
    - 4\alpha\beta + A^2 - \kappa A \bigg], \label{nc:accel}\\
 R &=& \frac{2}{a}\frac{d^2a}{d\tau^2} = \kappa \theta^2 \exp\left[
    \frac{2}{\kappa} (A - B) + 
    \frac{4}{\kappa} \beta e^{\kappa\theta t}\right]
    \bigg[\kappa \frac{A^2 - 4\alpha\beta}{\alpha
    e^{-\kappa\theta t} + \beta e^{\kappa\theta t} + A}
    \nonumber \\ 
 & & \qquad\qquad - (2\beta e^{\kappa\theta t} + A)^2
    - 4\alpha\beta + A^2 - \kappa A \bigg], \label{nc:R}
\end{eqnarray}
respectively. 

\begin{figure}[pt]
  \includegraphics{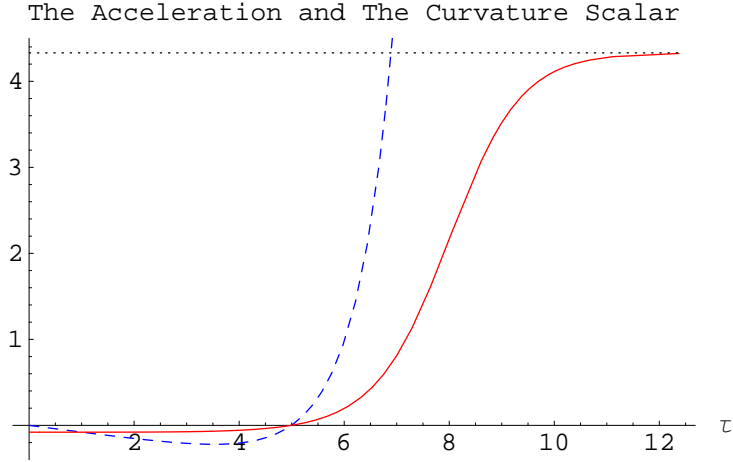}
  \caption{\label{fig:R} The solid line and the dashed line denote the
    curvature scalar and the acceleration of the scale factor,
    respectively. The dotted line is an asymptotic value of the
    curvature scalar as $\tau$ goes to infinity. Note that the
    comoving time is defined by $\tau > 0$. The curvature scalar comes
    to be a negative constant around $\tau=0$ and a positive constant
    as $\tau$ goes to infinity. This fact indicates that there is the
    phase transition from anti-de Sitter universe to de Sitter
    universe. This figure is plotted in the case of $\alpha=1$, $\beta
    = 1$, $A = -2$, $B = 1$, $ \kappa=1$, and $\theta = 4 $ in this
    BPP model.} 
\end{figure}

In order to describe the smooth transition from the decelerated phase
to the accelerated universe, eventually, from the AdS to the DS phase,
we will consider the special case of $\alpha>0$, $\beta>0$, and $A<
-\kappa$ with the condition $A^2 = 4 \alpha \beta$ in what follows. 
These constants tells us that the range of the conformal time is $t <
(\kappa\theta)^{-1} \ln (- A / 2\beta)$ as seen from
Eq. (\ref{nc:Omega}), and then the range of the comoving time should
be $\tau > 0$. Under this restriction, the expanding velocity
$da/d\tau$ is always positive and the scale factor increases from zero
to infinity. Note that $\tau(t)$ is a monotonic increasing function
with respect to $t$. On the other hand, the acceleration $d^2a/d\tau^2$ is
zero at the initial time $\tau=0$ and is negative before $\tau
=\tau_1$, where $\tau_1 = \int_{-\infty}^{t_1} e^{\rho(t)} dt$ where
$t_1 = (\kappa\theta)^{-1} \ln[(2\beta)^{-1} (-A - \sqrt{-\kappa
  A})]$. After $\tau=\tau_1$, the acceleration becomes positive,
which shows the smooth phase transition. Although the acceleration
diverges as $\tau$ goes to infinity, but there exists no curvature
singularity as shown in Fig.~\ref{fig:R} due to the infinite scale
factor. In fact, the curvature scalar is almost negative constant, $R
\approx -A(\kappa + A) \kappa\theta^2 \exp [\frac{2}{\kappa}(A - B)] <
0$, around $\tau = 0$ and it becomes zero at $\tau=\tau_1$, and then
approaches the positive constant, $R \approx -A \kappa^2\theta^2
\exp(-\frac{2}{\kappa} B) > 0$, at $\tau \rightarrow \infty$. This
fact shows that the phase transition from AdS universe to DS appears.

\begin{figure}
\includegraphics{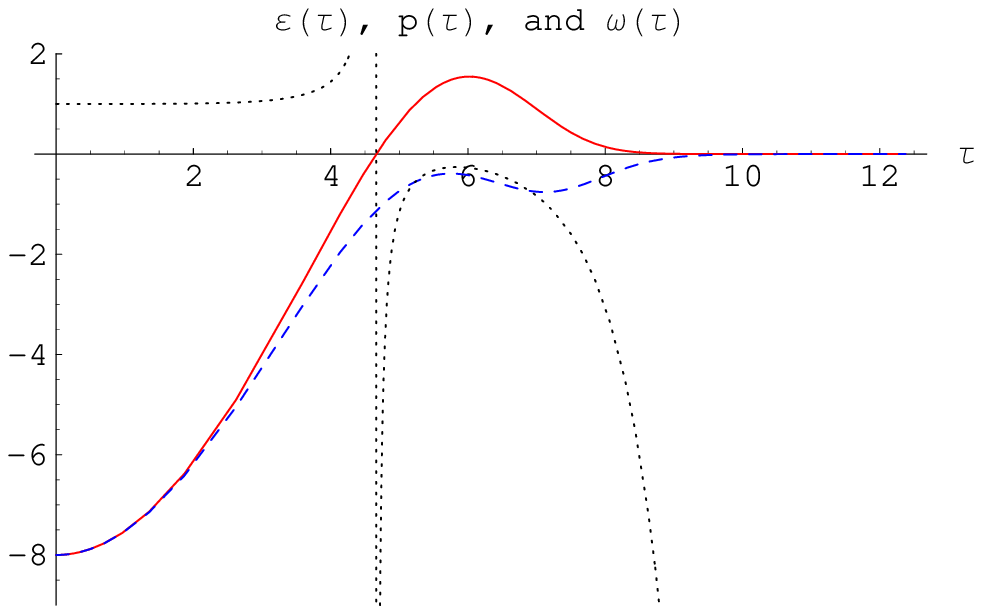}
\caption{\label{fig:energy} 
The solid, the dashed, and the dotted lines denote the energy density,
the pressure, and the state parameter of perfect fluid. Note that the
pressure is always negative, so that the state parameter can be
exotic. This figure is plotted with the same constants used in 
Fig.~\ref{fig:R}.}
\end{figure}


Now, the solutions (\ref{nc:Omega}) and ( \ref{nc:chi}) should  be
satisfied with the constraint (\ref{con}), which determines the
integration function $\kappa t_\pm$, 
\begin{equation}
\kappa t_\pm = \frac14 \kappa\theta^2 \left[ \kappa(\alpha
   e^{-\kappa\theta t} - \beta e^{\kappa\theta t}) -
   4\alpha\beta \right]. \label{nc:t}
\end{equation}
Then, the induced energy-momentum tensors (\ref{T++}), and (\ref{T+-})
are obtained as
\begin{eqnarray}
 T_{\pm\pm}^{\rm qt} &=& -\frac14 \kappa\theta^2 \left[
   \kappa(\alpha e^{-\kappa\theta t} + \beta e^{\kappa\theta t})
   - 4\beta e^{\kappa\theta t} (\alpha e^{-\kappa\theta t} -
   \beta e^{\kappa\theta t}) \right], \label{nc:T++} \\
 T_{+-}^{\rm qt} &=& \frac12 \beta \kappa^2\theta^2
   e^{\kappa\theta t}. \label{nc:T+-} 
\end{eqnarray}
Using Eqs.~(\ref{def:energy}) and (\ref{def:pressure}), the energy
density, the pressure are explicitly given as
\begin{eqnarray}
 \varepsilon &=& -\frac12 \kappa^2\theta^2 e^{2\kappa\theta t} (\alpha
    e^{-\kappa\theta t} + \beta e^{\kappa\theta t} + A) (\alpha
    e^{-\kappa\theta t} - \beta e^{\kappa\theta t}) \left(1 -
    \frac{4}{\kappa}\beta e^{\kappa\theta t} \right),
    \label{nc:energy} \\
 p &=& -\frac12 \kappa^2\theta^2 e^{2\kappa\theta t} (\alpha
    e^{-\kappa\theta t} + \beta e^{\kappa\theta t} + A)
    \bigg[ \alpha e^{-\kappa\theta t} + 3\beta e^{\kappa\theta
    t}  \nonumber \\
  & & - \frac{4}{\kappa} e^{\kappa\theta t} ( \alpha
    e^{-\kappa\theta t} - \beta e^{\kappa\theta t}) \bigg],
    \label{nc:p} 
\end{eqnarray}
so that the state parameter $\omega(\tau(t))$ reads
\begin{equation}
 \omega = \frac{\alpha e^{-\kappa\theta t} + 3\beta e^{\kappa\theta
    t} - \frac{4}{\kappa} e^{\kappa\theta t} ( \alpha
    e^{-\kappa\theta t} - \beta e^{\kappa\theta t})}{(\alpha
    e^{-\kappa\theta t} - \beta e^{\kappa\theta t}) \left(1 -
    \frac{4}{\kappa}\beta e^{\kappa\theta t} \right)} \label{nc:w}
\end{equation}
where its profile is plotted in Fig.~\ref{fig:energy} for the special
case giving the AdS-DS transition. The energy 
density and the pressure are the same value of $-1/2 \alpha^2
\kappa^2\theta^2$ approximately at the initial time $\tau=0$
corresponding to $t \rightarrow -\infty$, and then the state parameter
becomes $\omega\approx 1$. The energy density becomes zero at the
comoving time $\tau=\tau_2$, where $\tau_2 = \int_{-\infty}^{t_2}
e^{\rho(t)} dt$ where $t_2 \equiv [1/(\kappa\theta)] \ln
[\kappa/(4\beta)]$. It changes from negative value to positive around
$\tau=\tau_2$, but the pressure is always negative. The state
parameter diverges at $\tau_2$ since the energy density vanishes
faster than the pressure. The decelerated expansion of the early
universe is due to the negative energy density with the negative
pressure induced by quantum back reaction $(\omega > 0)$, and the
accelerated late-time universe comes from the positive energy and the
negative pressure which behave like dark energy source $(\omega < 0)$.

\section{Discussion}
\label{sec:dis}

We have shown that the phase changing transition from the AdS to the
DS phase is possible by assuming the modified Poisson brackets to the
semiclassical equations of motion in the BPP model. The usual BPP
model does not generate this kind of transition since the integration
function $t_\pm$ related to the vacuum state is trivially constant,
and the equation of state parameter is simply one which is independent
of the time. So, we have taken the nontrivial Poisson brackets at the
semiclassical level to overcome this triviality. 

The modified Poisson brackets are not the quantum commutators so that
it does not mean the quantization of the quantization since the fields
$\Omega$ and $\chi$ are not the operators. In fact, the modified
Poisson brackets can be applied to any stage of quantization in order
to modify the original equations of motion. For example, if one
considers the modified Poisson brackets at the classical dilaton
gravity, then the corresponding solution can be obtained, however, it
is difficult to obtain the meaningful 
solution in spite of its complexity. The other heuristic example may
be a two-dimensional simple harmonic oscillator with the mass $m$ and
the spring constant $k$, where its Hamiltonian is like
$H=(p_x^2+p_y^2)/(2m) + k(x^2 + y^2)/2$. The conventional Poisson
brackets generate the two independent set of Hamiltonian equations of
motion and then the well-known harmonic solutions are obtained. On the
other hand, at this classical level, if we assume the modified Poisson
brackets, $\{ x,p_x \}_{\rm MPB} = \{ y,p_y \}_{\rm MPB} = 1$, $\{
p_x, p_y \}_{\rm MPB} = \theta$, then the Hamiltonian equations of
motion are modified and the equations of motion can be written in the
second order form of $\ddot x + \omega^2 x = 2a \dot y$, $\ddot y +
\omega^2 y = -2a \dot x$, where $\omega = \sqrt{k/m}$ and $a =
\theta/(2m)$. The first order of Hamiltonian equations of motion have
been written in the form of the second order Euler-Lagrange equations
of motion in order to show the
explicit difference between the noncommutative case and the
commutative case. Then, the solutions are $x = x_0  \cos at \cos
(\omega' t + \varphi_1) + y_0 \sin at \cos(\omega' t + \varphi_2)$, $y
= y_0 \cos at \cos (\omega' t + \varphi_2) - x_0 \sin at \cos(\omega'
t + \varphi_1)$, where $\omega' \equiv \sqrt{\omega^2 + a^2}$ and
$x_0$, $y_0$, $\varphi_1$, and $\varphi_2$ are constants of
integration. Note that these are just modified classical solutions
rather than the quantum-mechanical ones.

The equation of state parameter is singular at a certain time as seen
in Fig.~\ref{fig:energy}. In order for the phase transition from the
ADS ($\omega >0$) to the DS universe $(\omega <0)$, the state parameter
also changes its signature at a certain time, in our case at $\tau =
\tau_2$. In fact, there are two options satisfying this condition. If
the energy density is always positive then the pressure should change
its sign, however, in this model, the pressure is always negative, so
that the energy density should change its sign. The latter case gives
the singular behavior. Of course, the quantum-mechanically induced
energy density allows the negative value.

One might wonder how to derive the nontrivial Poisson brackets which
are similar to the noncommutativity in string theory \cite{sw}. In
the string theory, the noncommutative brackets between the
coordinates are derived in the D-brane system applied in the constant
external tensor field. This is a higher dimensional realization of the
slowly moving point particle on the constant magnetic field. All of
these systems can be interpreted as constraint systems \cite{ko}, so
we can expect our model may be a similar constraint system, however,
it remains unsolved.


\begin{acknowledgments}
This work was supported by the Science Research Center Program of
the Korea Science and Engineering Foundation through the Center for
Quantum Spacetime \textbf{(CQUeST)} of Sogang University with grant
number R11 - 2005 - 021.
\end{acknowledgments}



\begin{thebibliography}{99}

\bibitem{carroll} S.\ M.\ Carroll, A.\ D.\ Felice, V.\ D., D.\ A.\ Easson,
  M.\ Trodden, and M.\ S.\ Turner, Phys.\ Rev.\ D \textbf{71}, 063513
  (2005). 

\bibitem{gardner} C.\ L.\ Gardner, Nucl.\ Phys.\ B \textbf{707}, 278
  (2005). 

\bibitem{cghs} C.\ G.\ Callan, S.\ B.\ Giddings, J.\ A.\ Harvey, and
  A.\ Strominger, Phys.\ Rev.\ D \textbf{45}, 1005 (1992).

\bibitem{bc} A.\ Bilal and C.\ Callan, Nucl.\ Phys.\ B \textbf{394},
  73 (1993).

\bibitem{rst}   J.\ G.\ Russo, L.\ Susskind, and L.\ Thorlacius,
  Phys.\ Rev.\ D \textbf{46}, 3444 (1992).

\bibitem{bpp} S.\ K.\ Bose, L.\ Parker, and Y.\ Peleg,
  Phys.\ Rev.\ Lett.\ \textbf{76}, 861 (1996).

\bibitem{bk} S.\ K.\ Bose and S.\ Kar, Phys.\ Rev.\ D \textbf{56}, 4444
  (1997). 

\bibitem{kv} W.\ Kummer and D.\ Vassilevich, Phys.\ Rev.\ D \textbf{60},
  084021 (1999).

\bibitem{ky:dg} W.\ T.\ Kim and M.\ S.\ Yoon, Phys.\ Lett.\ B
  \textbf{423}, 231 (1998).

\bibitem{rey}   S.\ J.\ Rey, Phys.\ Rev.\ Lett.\ \textbf{77}, 1929
  (1996). 

\bibitem{gv}  M.\ Gasperini and G.\ Veneziano, Phys.\ Lett.\ B
  \textbf{387}, 715 (1996).

\bibitem{ky:bd} W.\ T.\ Kim and M.\ S.\ Yoon, Phys.\ Rev.\ D
  \textbf{58}, 084014 (1998).

\bibitem{cdkz} M.\ H.\ Christmann, F.\ P.\ Devecchi, G.\ M.\ Kremer,
  and C.\ M.\ Zanetti, Europhys.\ Lett.\ \textbf{67}, 728 (2004).
 
\bibitem{ky:au} W.\ Kim and M.\ S.\ Yoon, Phys.\ Lett.\ B \textbf{645},
  82 (2007).

\bibitem{bn} G.\ D.\ Barbosa and N.\ Pinto-Neto, Phys.\ Rev.\ D
  \textbf{70}, 103512 (2004).

\bibitem{vassilevich} D.\ Vassilevich, \textit{Stability of a
    noncommutative Jackiw-Teitelboim gravity}, hep-th/0602095.

\bibitem{ko} W.\ T.\ Kim and J.\ J.\ Oh, Mod.\ Phys.\ Lett.\ A
  \textbf{15}, 1597 (2000).

\bibitem{bk:H} A.\ Bilal and I.\ I.\ Kogan, Phys.\ Rev.\ D \textbf{47},
  5408 (1993).

\bibitem{sw} N.\ Seiberg and E.\ Witten, J.\ High Energy
  Phys.\ \textbf{09}, 032 (1999).

\end{thebibliography}
\end{document}